\documentstyle[12pt,epsf]{article}

\newcommand{\ol}{\overline}
\begin{document}

\baselineskip=18pt plus 0.2pt minus 0.1pt

\begin{titlepage}
\title{
\hfill\parbox{4cm}
{\normalsize KUNS-1494\\HE(TH)~98/02\\{\tt hep-th/9802189}}\\
\vspace{3cm}
$E_8$ flavour multiplets
}
\author{
Yosuke Imamura\thanks{{\tt imamura@gauge.scphys.kyoto-u.ac.jp}}
{}\thanks{
Supported in part by Grant-in-Aid for Scientific
Research from Ministry of Education, Science and Culture
(\#5416).}
\\[7pt]
{\it Department of Physics, Kyoto University, Kyoto 606-01, Japan}
}
\date{}

\maketitle
\thispagestyle{empty}

\begin{abstract}
\normalsize
We analyze gauge symmetry enhancements
$SO(16)\rightarrow E_8$ on eight D7-branes
and $SO(14)\times U(1)\rightarrow E_8$ on seven D7-branes
from open strings.
String configurations which we present in this paper are
closely related to the ones given by
Gaberdiel and Zwiebach.
Our construction is based on $SO(8)\times SO(8)$
decomposition and its relation to the D8-brane case
via T-duality is clearer.
Then we study supersymmetric Yang-Mills theory on D3-brane near the D7-branes.
This theory has flavour symmetry group which is equal to the gauge group
on D7-branes.
We suggest that when this symmetry is enhanced, two dyons
make bound states which, together with elementary quarks,
constitute an $E_8$ multiplet.
\end{abstract}

\end{titlepage}

\section{Introduction}
In recent years, string theory has uncovered
non-perturbative properties of Yang-Mills theories one after another.
One of these interesting results
is the existence of non-trivial fixed points where
exceptional flavour symmetries are realized.

The only string theory in which the appearance of exceptional gauge groups
can be analyzed perturbatively is the heterotic string theory.
By using dualities connecting different string theories,
we can conclude that the enhancement occurs in other string theories
on special points in their moduli spaces.
Although dualities are convenient tools,
it does not cover the whole moduli space.
Therefore, it is important to investigate the mechanism
of gauge symmetry enhancement in each theory.

In refs.\cite{Seiberg,MorrisonSeiberg},
Yang-Mills theories on D4-branes in type-I' theory with particular D8-branes
background are analyzed.
The exceptional group arises as gauge symmetry on D8-branes.
Type-I' theory contains $16$ D8-branes.
If $n$ of them coincide, $U(n)$ gauge symmetry arises.
If these $n$ D8-branes coincide with an orientifold 8-plane,
the symmetry is enhanced to $SO(2n)$ perturbatively.
Furthermore, if string coupling constant on the orientifold plane diverges,
D-particles on the orientifold plane become massless and
the symmetry is enhanced to exceptional group $E_{n+1}$
non-perturbatively\cite{Evidence,Creation}.

On the other hand, in $S^2$ compactified type-IIB theory
which is equivalent to K3 compactification of F-theory,
exceptional gauge groups arise
when particular 7-branes coincide.
In a region of moduli space where 24 7-branes are regarded
as four orientifold 7-planes and sixteen D7-branes,
this theory can be understood as T-dual of type-I' theory.
Two of the four orientifold 7-planes come
from one orientifold 8-plane in type-I' theory,
and the other two from another orientifold plane.
The condition that string coupling constant on one orientifold 8-plane diverges
is transformed into the coincidence of two orientifold 7-planes.
In terms of type-IIB theory, all modes in $\bf248$ multiplet of $E_8$
can be understood as open string configurations\cite{FromOpen}.
One of the purposes of this paper is to investigate
the relation of these two viewpoints
of gauge symmetry enhancement phenomena.

By using gauge symmetry enhancement on D7-branes,
we can construct field theories with exceptional group flavor symmetries
as theories on D3-branes.
The field theory on $k$ D3-branes in type-IIB theory compactified
on orientifold $T^2/Z_2$
is supersymmetric (${\cal N}=2$) $Sp(k)$ Yang-Mills theory with $16$
flavors.
(In this paper, we focus only on $Sp(1)=SU(2)$ gauge theory.)
These $16$ hypermultiplets come from open strings
stretched between the D7-branes and the D3-branes.
When global symmetry is enhanced to $E_{n+1}$, many extra degrees of freedom
should appear.
For example,
if $SO(14)\times U(1)$ is enhanced to $E_8$,
the quark multiplet belonging to $\bf14$ is enlarged to, at least,
the $\bf248$ representation
and $248-14$ extra degrees of freedom are necessary.
The second purpose of this paper is to show
what these extra degrees are.

\section{Gauge symmetry enhancement}

\begin{figure}[hbt]
\centerline{
\epsfxsize=12cm
\epsfbox{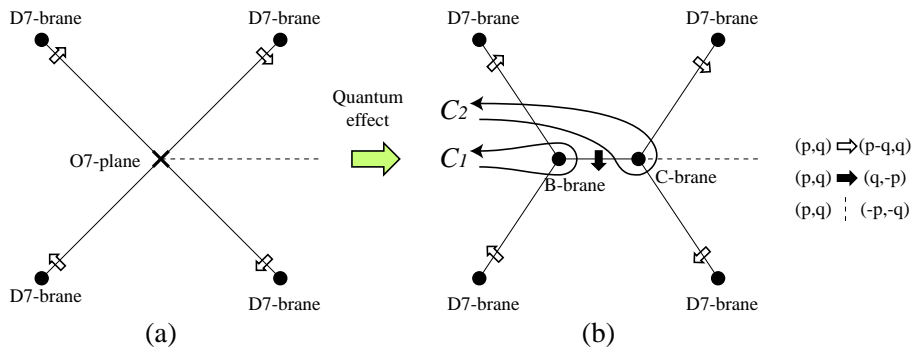}
}
\caption{(a) One block contains four D7-branes (blobs) and an orientifold plane
(small cross at the center).
Branch cuts, which gives transformation $(p,q)\rightarrow (p-q,q)$,
stretched between each D7-brane and the orientifold plane (solid lines).
Furthermore, a branch cut reversing the string orientation goes
from the orientifold plane to infinity (broken line).
(b) Quantum effects split an orientifold plane into two 7-branes.
We call them B- and C-brane, and
the monodromies along paths $C_1$ and $C_2$ are $(p,q)\rightarrow(p,p+q)$ and
$(p,q)\rightarrow(3p-4q,p-q)$, respectively.}
\label{fig:fwq}
\end{figure}

As we have mentioned,
one of the purposes of this paper
is to clarify the relationship between gauge symmetry enhancements
on D7-branes\cite{FromOpen} and those on D8-branes\cite{Creation}.
These theories are related by T-duality, and it is best understood
in the case of flat background, in which the charge of each orientifold plane
is canceled by D-branes.
Therefore it is convenient to make up blocks containing
one orientifold plane and D-branes
whose charges cancel out each other.
In the case of type-IIB on $T^2/Z_2$, each block contains
one orientifold 7-plane and four D7-branes((a) in Fig.\ref{fig:fwq}).
Branch cuts, which gives transformation $(p,q)\rightarrow (p-q,q)$,
stretched between each D7-brane and the orientifold plane.
Furthermore, a branch cut reversing the string orientation goes
from the orientifold plane to infinity.
This corresponds to orientifold flip.
If all these 7-branes coincide, $SO(8)$ gauge symmetry realizes on them.
In this section, we construct string configurations
belonging to several representations of $SO(8)$.
Then, we will combine them to make $\bf248$ representation of $E_8$.

\subsection{Adjoint ($\bf28$) representation}
\begin{figure}[hbt]
\centerline{
\epsfbox{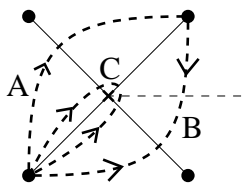}
}
\caption{Open strings belonging to adjoint representation of $SO(8)$.
The blobs and the small cross represent D7-branes and an orientifold plane,
respectively.
A and B should be distinguished. C is projected out by
orientifold projection.}
\label{fig:so8adj}
\end{figure}

States contained in the adjoint $\bf28$ representation of $SO(8)$
are generated by open strings stretched between two D7-branes in a block.
Because strings can not pass through the orientifold planes
without creating new string by Hanany-Witten effect\cite{HananyWitten},
we should distinguish strings which pass the different sides
of an orientifold plane.
The strings are represented as A and B in Fig.\,\ref{fig:so8adj}.
(Because strings winding around an orientifold plane twice are removable,
winding number around an orientifold plane is defined by mod $2$.)
In this paper, when string charge is not specified explicitly,
broken lines represent fundamental strings,
while solid lines represent strings with D-string charge.
Strings connecting a D7-brane and its mirror
(C in Fig.\,\ref{fig:so8adj})
are not allowed by the orientifold projection.
As a result, we obtain $28$ states altogether.

\subsection{Vector (${\bf8}_v$) representation}
\begin{figure}[hbt]
\centerline{
\epsfxsize=15cm
\epsfbox{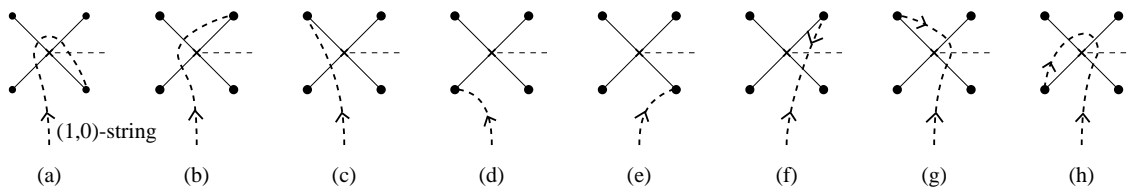}
}
\caption{String configurations belonging to vector representation of $SO(8)$}
\label{fig:so8vec}
\end{figure}

Vector representation is constructed with
open strings one of whose end points is on a D7-brane in a block.
We should distinguish two configurations
in which strings pass through different sides of an orientifold plane.
There are eight different configurations as shown in Fig\,\ref{fig:so8vec}.
\subsection{Spinor (${\bf8}_s$ and ${\bf8}_c$) representations
and decouplings of D7-branes}
In \cite{FandO}, it is pointed out that $SL(2,{\bf Z})$ duality
transformation of type-IIB theory causes automorphism
of $SO(8)$ gauge group on orientifold 7-planes.
By means of this automorphism, we can easily construct spinor representations
${\bf8}_s$ and ${\bf8}_c$ from vector representation ${\bf8}_v$.
For this purpose, it is necessary to take account of quantum effects
which split
an orientifold planes into two 7-branes\cite{FandO} ((b) in Fig.\ref{fig:fwq}).
Following ref.\cite{FromOpen}, we call them B- and C-brane.
This splitting of orientifold planes is the same phenomenon as
the appearance of two singularities (monopole and dyon singularities)
on the $u$-plane of ${\cal N}=2$ $SU(2)$ supersymmetric Yang-Mills theory,
and we can study this configuration by means of, for example,
Seiberg-Witten curve for $N_f=4$ $SU(2)$ supersymmetric Yang-Mills theory
\cite{SW}.
It is known that the monodromies
along paths $C_1$ and $C_2$ in (b) of Fig.\ref{fig:fwq}
are $(p,q)\rightarrow (p,p+q)$ and $(p,q)\rightarrow(3p-4q,p-q)$, respectively.
To realize these monodromies, new branch cut is necessary
between the B-brane and the C-brane.
If strings cross the cut downward, their charge $(p,q)$ is transformed
into $(q,-p)$.
Of cause, the transformation depends
upon the way to attach the cuts stretched from
D7-branes and infinity on the B-brane and C-brane.
In our convention, two of four cuts from D7-branes are attached on
each of the B- and C-brane and the cut reversing string orientation
goes from C-brane to infinity as is shown in (b) of Fig.\ref{fig:fwq}.

\begin{figure}[hbt]
\centerline{
\epsfxsize=9cm
\epsfbox{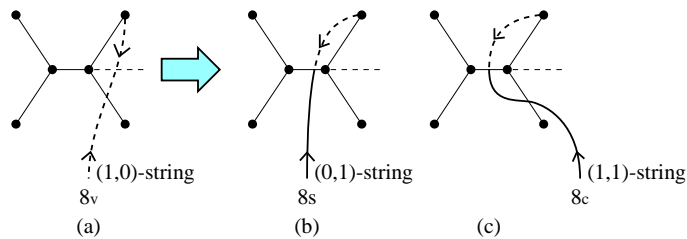}
}
\caption{From string configurations in ${\bf8}_v$ representation,
configurations belonging to ${\bf8}_s$ (b)
and ${\bf8}_c$ (c) representations can be constructed.}
\label{fig:so8spn}
\end{figure}

\begin{figure}[hbt]
\centerline{
\epsfbox{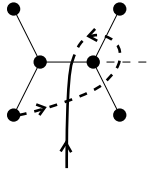}
}
\caption{Configurations which have self intersections which
are not able to be removed by continuous deformation
should be forbidden.}
\label{fig:forbidden}
\end{figure}

\begin{figure}[hbt]
\centerline{
\epsfxsize=15cm
\epsfbox{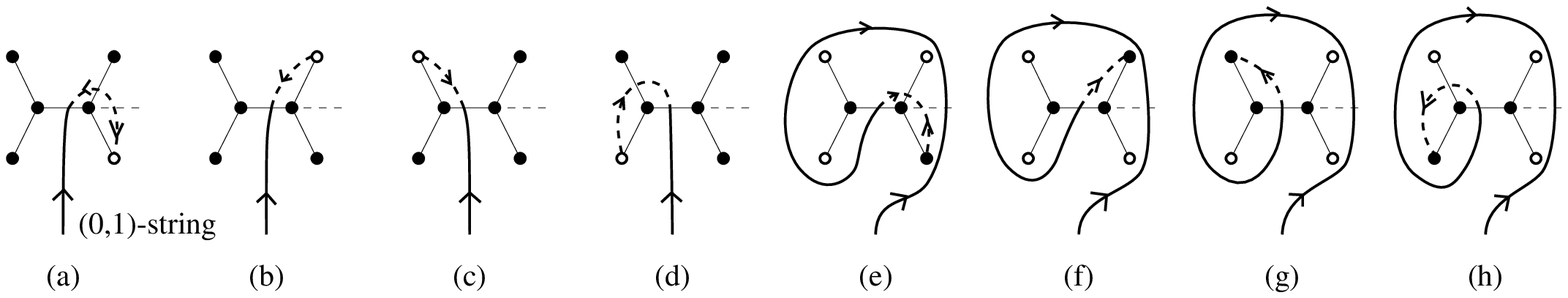}
}
\caption{String configurations belonging to ${\bf8}_s$ representation.}
\label{fig:8sconfig}
\end{figure}

By means of the connection between $SL(2,{\bf Z})$ transformation
and the automorphism of $SO(8)$ gauge symmetry,
we can easily make configurations belonging to the spinor representations.
The branch cut going between the two 7-branes generated by the splitting of
orientifold plane causes the S-dual transformation
$\tau\rightarrow-1/\tau$, and the corresponding automorphism swaps
vector and spinor representations.
Therefore, we get ${\bf8}_s$ representation
by letting strings in ${\bf8}_v$ representation ((a) of Fig.\,\ref{fig:so8spn})
pass through this branch cut
((b) of Fig.\,\ref{fig:so8spn}).
As a result, the strings are converted to $(0,1)$-strings.
Applying this operation naively,
it might seem that a configuration
in Fig.\,\ref{fig:forbidden} is obtained.
However, we should forbid such configurations, in which a string
crosses with itself,
so that decoupling argument given below works well.
This rule may be understood by
investigating these configurations in M-theory picture.
However, we do not discuss about this point in this paper
and we assume the selection rule%
\footnote{After this paper was submitted, the works \cite{Geo,BPScond} appear
which explain that the selection rule is obtained
from BPS condition for strings.}.
Taking account of this rule, we obtain eight independent configurations
in Fig.\,\ref{fig:8sconfig}.
As we have already mentioned,
we represent fundamental strings by dashed lines
and strings which have nonzero D-string charge by solid lines.
Arrows on solid lines represent the orientation with respect to D-string charge.

\begin{figure}[hbt]
\centerline{
\epsfbox{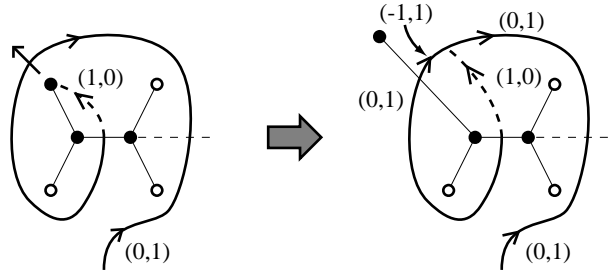}
}
\caption{Even if the upper-left D7-brane,
on which a fundamental string is attached,
is removed, this string configuration does not get massive
due to Hanany-Witten effect.}
\label{fig:decouple}
\end{figure}

To confirm that these configurations belong to spinor representations,
let us consider the decoupling properties of them.
In Fig.\,\ref{fig:8sconfig}, the D7-branes
whose decoupling makes the states massive,
are represented by circles and the others by blobs.
(We assume that the decoupled D7-branes move straight outward.)
If we remove one of the four D7-brane away,
four of the eight configurations remain light.
For example,
if we decouple the upper-left D7-brane,
it is clear that (a), (b) and (d) in Fig.\,\ref{fig:8sconfig} remain
as light states.
In addition to them, (g) in Fig.\,\ref{fig:8sconfig} also stays light.
If the D7-brane is decoupled, a fundamental string with the opposite orientation
of the original fundamental string is generated due to the Hanany-Witten effect,
and the string, which was attached on the decoupled D7-brane,
is reconnected to D-string (Fig.\,\ref{fig:decouple}).
As a result, we get a branching rule ${\bf8}_s\rightarrow{\bf4}+\ol{\bf4}$
($\ol{\bf4}$ becomes massive and decouples).
By every further decoupling procedure of D7-branes,
the number of light states is halved. This is what is required for
spinor representations.

\begin{figure}[hbt]
\centerline{
\epsfxsize=15cm
\epsfbox{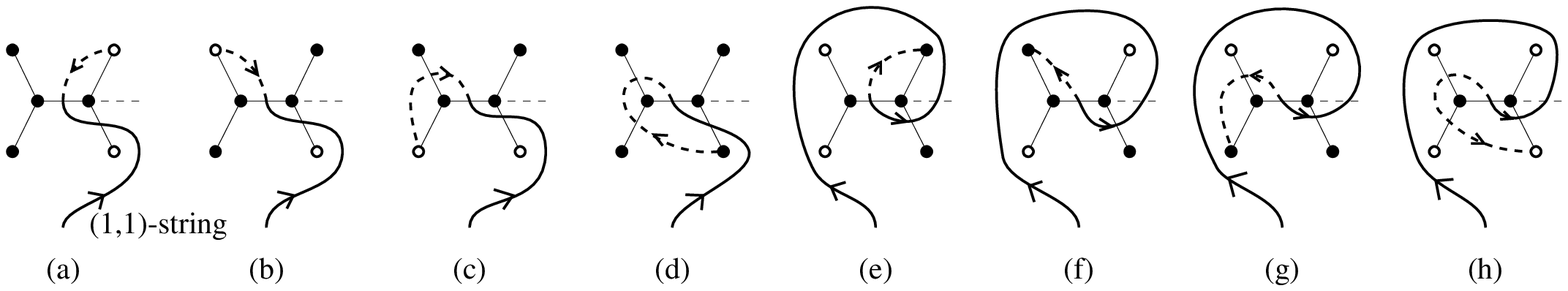}
}
\caption{String configurations belonging to ${\bf8}_c$ representation.}
\label{fig:8cconfig}
\end{figure}

Configurations belonging to ${\bf8}_c$ representation are also
constructed by making the strings in ${\bf8}_v$ representation
pass through the branch cuts
as in (c) of Fig.\,\ref{fig:so8spn}. In this case, the strings are converted to
$(1,1)$-strings.
Taking the selection rule into account,
we get eight configurations in Fig.\,\ref{fig:8cconfig}

If we attach these configurations to a D3-brane, we obtain monopoles and dyons
of $SU(2)$ gauge theory on the D3-brane.
They belong to fundamental representation $\bf2$ of $SU(2)$ gauge group.
Each configuration in Fig.\,\ref{fig:8sconfig} and Fig.\,\ref{fig:8cconfig}
is one of the doublet.
Another states in the doublet can be
obtained by adding a D-string or $(1,1)$-string corresponding to a
generator of $SU(2)$,
which, starting from the D3-brane, goes around the block and returns
to the D3-brane. 
As a result, we get configurations
similar to the ones in Fig.\,\ref{fig:8sconfig} and
Fig.\,\ref{fig:8cconfig} with opposite string direction.

\subsection{$E_8$ gauge symmetry}
\begin{figure}[hbt]
\centerline{
\epsfbox{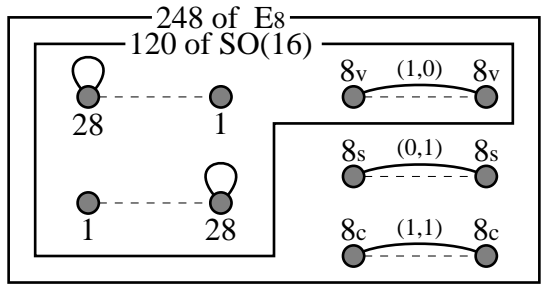}
}
\caption{Constituents of $\bf248$ representation of $E_8$.
Dark circles represent blocks containing six 7-branes respectively.}
\label{fig:e8byso8so8}
\end{figure}

As we mentioned in the Introduction,
when two blocks, each of which consists of an orientifold plane
and four D7-branes,
are overlapped at one point,
gauge symmetry is enhanced to $E_8$.
Adjoint representation $\bf248$ of $E_8$ is
decomposed to $({\bf28},{\bf1})$, $({\bf1},{\bf28})$,
$({\bf8}_v,{\bf8}_v)$, $({\bf8}_s,{\bf8}_s)$ and $({\bf8}_c,{\bf8}_c)$
of $SO(8)\times SO(8)\subset SO(16)\subset E_8$.
We represent these pairs by $(R_1,R_2)$.
$R_1$ and $R_2$ are $SO(8)$ representations,
which are realized on the left block and the right one respectively.
Combining $SO(8)$ configurations according to this decomposition,
we obtain the string configurations of $E_8$ gauge field
(Fig.\,\ref{fig:e8byso8so8}).
In these configurations, we connect two branch cuts
reversing the string orientation (Thin broke lines in Fig.\ref{fig:e8byso8so8}),
and the monodromy around the infinity is one.
Picking up states which does not contain D-string,
we obtain adjoint ($\bf120$) representation of $SO(16)\subset E_8$.
Therefore $E_8$ gauge symmetry is understood
as the $SL(2,{\bf Z})$ completion of
perturbative gauge group $SO(16)$.

At first sight, it seems strange that only $(1,0)$, $(0,1)$ and $(1,1)$ strings
appear in the above arguments.
These are not invariant under $SL(2,{\bf Z})$ transformation.
In fact, all $(p,q)$ strings contribute to the gauge symmetry enhancement,
because we can add string loops enclosing two blocks.
If we shrink the string loop,
it is reduced to two strings stretched between two blocks.
It may seem that these strings have opposite orientation
each other and their charges are canceled.
However, because a branch cut reversing string orientation goes
from one block to another,
we should regard these strings have the same orientation.
We define charge of string going between two blocks
so that string loops in clockwise orientation
give positive contribution to the charge.
For example, if we add a fundamental string wrapped $n$ times around
the whole two blocks in the clockwise orientation,
$(p,q)$ string going between the two blocks
changes to $(2n+p,q)$ strings.
Existence of these infinite tower of states means the appearance
of new spatial dimension
and we obtain decompactified theory on D8-brane in T-dual picture.
Similarly, ambiguity of D-string charge is understood
as the appearance of the eleventh direction of M-theory.
As a result, $E_8$ gauge field lives in ten dimensional space-time.
In this way, we should interpret $(1,0)$, $(0,1)$ and $(1,1)$ strings
in the above arguments
as $(2n+1,2m)$, $(2n,2m+1)$ and $(2n+1,2m+1)$ strings,
where $m$ and $n$ are arbitrary integers.
Each of the three sets of string charges is separately invariant under 
$SL(2,{\bf Z})$.
The correspondence between $SO(8)$ representations $R_1$ and the string charges
is given in the left column in Table \ref{tbl:branch}.

\subsection{$E_8$ gauge symmetry on seven D7-branes}
Until now, we have explained the enhancement of $SO(8)\times SO(8)$
gauge symmetry 
on eight D7-branes
to $E_8$.
Each $SO(8)$ factor is realized on a fixed point of $T_2/{\bf Z}_2$.
However, we cannot construct $SU(2)$ Yang-Mills theory
on D3-brane in the background,
because when $E_8$ symmetry is realized,
one cycle of the $T^2$ shrinks to a point.
In this case, a theory on D3-brane should be regarded as a field theory
on five-brane in M-theory.
Furthermore, quark multiplet,
which belongs to the vector (${\bf16}$) representation of $SO(16)$,
cannot be extended to $E_8$ multiplet,
because no $E_8$ representation contains vector representation of
$SO(16)$.
To avoid these difficulties, we should consider the configuration
where seven D7-branes are on the orientifold plane\cite{Seiberg}.
In this case, the largest perturbative gauge symmetry is $SO(14)\times U(1)$.
To realize this, we should remove one D7-brane from the orientifold.
At first sight, it seems that the decoupling of a D7-brane
causes gauge symmetry breaking.
However, as we will show below, gauge symmetry is not broken.
Instead, the dimension of the space where the gauge field lives
decreases.

\begin{figure}[hbt]
\centerline{
\epsfbox{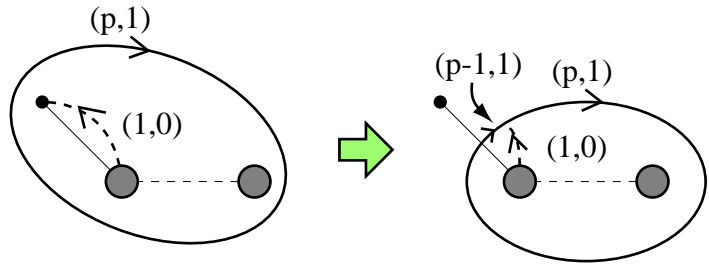}
}
\caption{Open strings attached to the decoupled D7-brane can be removed by
adding a loop of D-string enclosing the whole two blocks containing
the decoupled D7-brane.}
\label{fig:removestring}
\end{figure}

If we decouple one D7-brane in the right block,
gauge symmetry $SO(8)$ is broken to $SO(6)\times U(1)$
and the representation $R_1$ is decomposed into some representations
of $SO(6)\times U(1)$.
We represent them by $R'_1$.
First, let us focus on the adjoint representation $\bf28$ of $SO(8)$.
It is
 decomposed to ${\bf15}_0+{\bf6}_{+1}+{\bf6}_{-1}+{\bf1}_0$ of $SO(6)\times U(1)$.
\begin{figure}[hbt]
\centerline{
\epsfbox{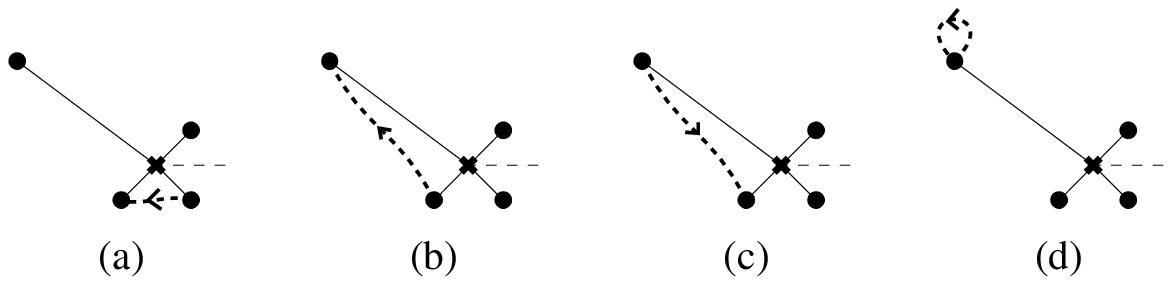}
}
\caption{Branching of
         ${\bf28}\rightarrow{\bf15}_0+{\bf6}_{+1}+{\bf6}_{-1}+{\bf1}_0$.}
\label{fig:28to}
\end{figure}
In this case, the number of strings
going into the decoupled D7-brane is equal to a $U(1)$ charge $Q_A(R'_1)$
of $R'_1$, where the normalization of $Q_A(R'_1)$ is defined so that
two $\bf6$ representations of $SO(6)$ come from $\bf28$ of $SO(8)$
have $Q_A=\pm1$.
When $U(1)$ charges does not vanish, the open string is attached on
removed D7-brane (Fig.\,\ref{fig:28to}).
Therefore, these states become massive and decouple and gauge group is broken to $SO(6)\times U(1)$.
As long as we consider one block, this is the all story for the decoupling
of a D7-brane.
However, if we use another block, we can restore the gauge symmetry. 
Remember that we can add a string loop enclosing the whole two blocks containing
the decoupled D7-brane.
If a configuration contains a string going into the decoupled D7-brane,
we can detach the string from the D7-brane
by adding a wrapping D-string in the clockwise orientation
and shrinking it (Fig.\,\ref{fig:removestring}).
When the loop passes through the decoupled D7-brane,
a fundamental string in opposite direction of the original string is created
due to the Hanany-Witten effect,
and the strings are detached from decoupled D7-brane.
If a configuration contains a string going out from decoupled D7-brane,
we should add the D-string loop with anti-clockwise orientation.
Therefore, states in ${\bf 6}_{\pm1}$ representation do not decouple
and $E_8$ gauge symmetry is recovered.
However, the ambiguity of D-string charges is now fixed
and the $E_8$ gauge field live on nine dimensional space-time.
This is regarded as field theory on D8-brane in T-dual picture.
The vanishing of the M-theoretic direction
might be related to the difficulties of M-theoretic interpretation of
massive IIA theory.

\begin{figure}[hbt]
\centerline{
\epsfbox{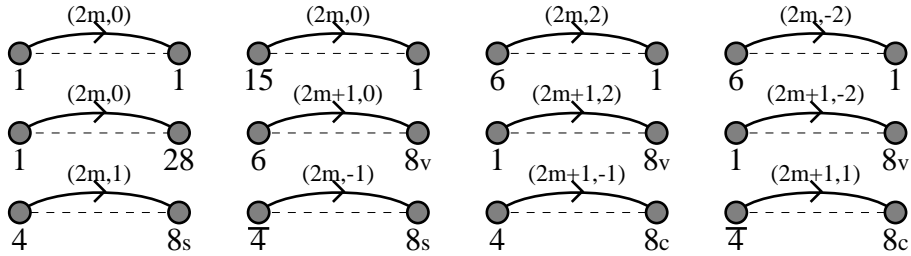}
}
\caption{String configurations in $\bf248$ representation of $E_8$
on seven D7-branes. Dark circles on the left represent blocks
containing three D7-branes and an orientifold plane.
(One D7-brane is decoupled.)
The right blocks contain four D7-branes and an orientifold planes.
String charge $p$ is defined by mod $2$.}
\label{fig:68}
\end{figure}

The decreasing of the dimension explained above
can be understood more clearly in the following way.
If we add a D-string loop winding $N$ times around the two blocks
in the clockwise orientation and shrink it, 
D-string charge $q$ between the two blocks
increases by $2N$, while
the number $n_F$ of strings going into the decoupled D7-brane decreases by $N$.
Therefore, linear combination $c=n_F+q/2$ is invariant when we add string loops.
\begin{figure}[hbt]
\centerline{\epsfbox{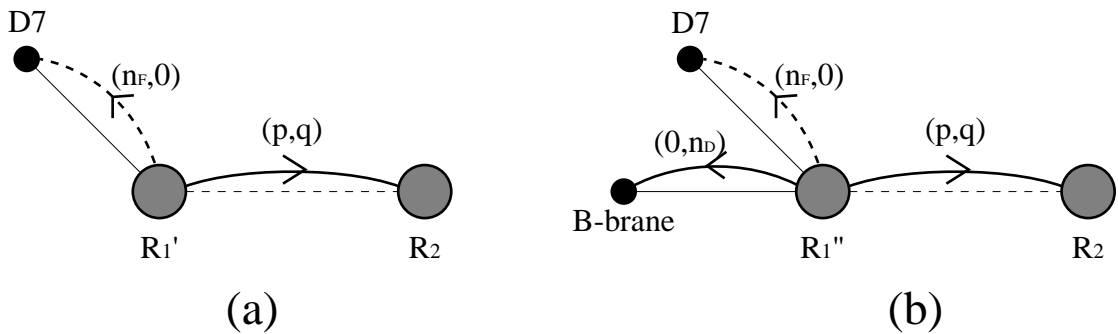}}
\caption{String configurations of nine dimensional $E_8$ gauge theory (a) and
ones of eight dimensional $E_8$ gauge theory (b).}
\label{fig:987}
\end{figure}
By using string configurations which we have already given,
we can check explicitly that the constant $c$
takes the same value for each configuration
belonging to the same representation of $SO(6)\times U(1)$,
and it is equal to $Q_A(R'_1)$.
Then, we obtain a equation
\begin{equation}
n_F+\frac{1}{2}q=Q_A(R'_1).
\label{reduction}
\end{equation}
If we require $n_F=0$, eq.\,(\ref{reduction}) restricts
the two dimensional $(p,q)$-charge lattice to a one dimensional lattice.
This restriction corresponds to the disappearance of the M-theoretic direction.
(Note that $p$ is not determined by eq.\,(\ref{reduction}) and its
ambiguity persists.) 
The middle column of table \ref{tbl:branch} shows
the correspondence between representation $R'_1$ of $SO(6)\times U(1)$ and
string charge $(p_2,q_2)$ between two blocks.
We can construct string configurations of $E_8$ gauge field by connecting
two blocks, which belong to $R'_1$ of $SO(6)$ and $R_2$ of $SO(8)$,
by $(p_2,q_2)$ strings, whose charge is given in Table \ref{tbl:branch}.
The result is shown in Fig.\,\ref{fig:68}.

\begin{figure}[hbt]
\centerline{
\epsfbox{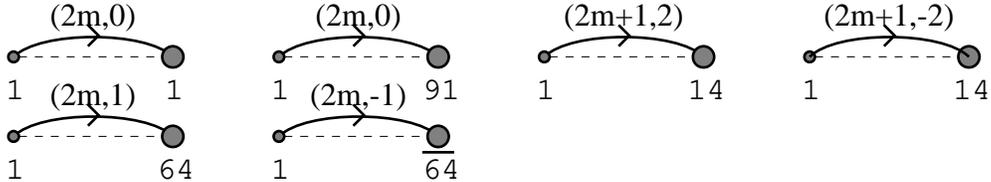}
}
\caption{String configurations in $\bf248$ multiplet.
Small circles contain an orientifold plane and large circles
contain seven D7-branes and an orientifold plane, respectively.}
\label{fig:so14}
\end{figure}

In order to make contact with the result of Type-I' theory\cite{Creation},
we should give the configurations where the largest perturbative symmetry
$SO(14)\times U(1)$ is realized.
They are obtained by gathering the seven D7-branes together at one
orientifold plane.
In spinor configurations, open string attached on
one of D7-branes and pass through between B-and C-brane in each blocks.
Therefore, if we move D7-branes in one block to another block,
the open string is hooked on the B- or C-brane and
string charge $p$ between the two blocks is shifted by one.
As a result, we obtain the configurations in Fig.\,\ref{fig:so14}.
Via T-duality, the D-string charge between two orientifold planes
is related to the D-particle charge in type-I' theory
and it corresponds to string winding number in heterotic string theory.
By this relation, these configurations completely coincide
with the spectrum given in \cite{Creation}.

\subsection{Further decoupling of a B-brane}

In the last subsection, we show that the decoupling of one D7-brane decreases
the spatial dimension from nine to eight.
If we want to construct $E_8$ gauge theory in seven spatial dimension,
we should decouple one B-brane, in addition to one D7-brane.
We have explained in the last subsection that
the reason of decreasing of the dimension is that string loops
with D-string charge are forbidden due to
branch cut stretched from decoupled D7-brane.
Further decoupling of a B-brane create a new branch cut giving transformation
$(p,q)\rightarrow(p,p+q)$, which is equivalent to the monodromy
along the path $C_1$ of (b) in Fig.\ref{fig:fwq}.
The composition of transformations
given by two branch cut from two decoupled branes
forbid string loops with any charge.
Therefore, enhancement of spatial dimension does not occur
and we obtain seven dimensional field theory.

We can determine string charge between two blocks
fixed by the decoupling of a B-brane in a similar way of the last subsection.
We have two options for a decoupled B-brane.
Here we suppose that we decouple a B-brane in the block
from which a D7-brane is removed.
For configurations in vector ($\bf6$) and adjoint ($\bf15$) representations
of $SO(6)$, states staying light in the decoupling of B-brane
is identical to ones in decoupling of
both B- and C-brane, (or orientifold 7-plane),
because no configuration in the representations
contain strings which pass through between B- and C-brane.
Therefore, by this decoupling,
gauge group $SO(6)$ is broken to $U(3)=SU(3)\times U(1)$.
Under this symmetry breaking,
representations $R'_1$
are decomposed into some representations of $SU(3)\times U(1)$.
We represent them by $R''_1$.
This branching is shown in Table \ref{tbl:branch}
(The middle and the right columns).
Furthermore, following the same argument in the last subsection,
we can deduce a relation
\begin{equation}
n_D+\frac{1}{p}=\frac{1}{2}[Q_B(R''_1)-Q_A(R''_1)],
\label{eq2}
\end{equation}
where $n_D$ is number of D-strings going into decoupled B-brane,
$Q_B$ is a charge under $U(1)$ factor in $SU(3)\times U(1)$
and $p$ is a fundamental string charge between two blocks
((b) in Fig.\ref{fig:987}).
(We determined the right hand side of eq.(\ref{eq2}) by
deforming each configuration which we have given explicitly.)
The $U(1)$ charge $Q_B$ is normalized so that representations coming from
spinor representation of $SO(6)$ have half integral charge.
Using eq.(\ref{reduction}) and eq.(\ref{eq2}),
fundamental and D-string charge are fixed. It is given in the right column
in Table \ref{tbl:branch} as $(p_3,q_3)$.
We can construct string configurations of seven dimensional $E_8$ gauge theory
by connecting two blocks belonging to representation $R''_1$ and $R_2$
by $(p_3,q_3)$-strings like (b) of Fig.\ref{fig:987}.
They are equivalent to ones given in ref.\cite{FromOpen}.

\begin{table}[hbt]
\caption{Branching of $SO(8)$ representation $R_1$
in symmetry breaking
$SO(8)\rightarrow SO(6)\times U(1)\rightarrow SU(3)\times U(1)\times U(1)$.}
\label{tbl:branch}
\centerline{%
\begin{tabular}{cc|cc|cc}
\hline
$R_1$ & $(p_1,q_1)$ &
$R'_1$ & $(p_2,q_2)$ &
$R''_1$ & $(p_3,q_3)$ \\
\hline\hline
$28$ & $(2m,2n)$ & $15_0$ & $(2m,0)$ & $8_{0,0}$ & $(0,0)$ \\
     &           &        &          & $3_{0,+2}$ & $(+2,0)$ \\
     &           &        &          & $\ol3_{0,-2}$ & $(-2,0)$ \\
     &           &        &          & $1_{0,0}$ & $(0,0)$ \\
\cline{3-6}
     &           & $6_{+1}$ & $(2m,+2)$ & $3_{+1,-1}$    & $(-2,+2)$ \\
     &           &          &           & $\ol3_{+1,+1}$ & $(0,+2)$ \\
\cline{3-6}
     &           & $6_{-1}$ & $(2m,-2)$ & $3_{-1,-1}$    & $(0,-2)$ \\
     &           &          &           & $\ol3_{-1,+1}$ & $(+2,-2)$ \\
\cline{3-6}
     &           & $1_0$    & $(2m,0)$  & $1_{0,0}$      & $(0,0)$ \\
\hline
$8_v$ & $(2m+1,2n)$ & $6_0$ & $(2m+1,0)$ & $3_{0,-1}$    & $(-1,0)$ \\
      &             &       &            & $\ol3_{0,+1}$ & $(+1,0)$ \\
\cline{3-6}
      &             & $1_{+1}$ & $(2m+1,+2)$ & $1_{+1,0}$ & $(-1,+2)$ \\
\cline{3-6}
      &             & $1_{-1}$ & $(2m+1,-2)$ & $1_{-1,0}$ & $(+1,-2)$ \\
\hline
$8_s$ & $(2m,2n+1)$ & $4_{+1/2}$ & $(2m,+1)$ & $3_{+1/2,+1/2}$ & $(0,+1)$ \\
      &             &            &           & $1_{+1/2,-3/2}$ & $(-2,+1)$ \\
\cline{3-6}
      &          & $\ol4_{-1/2}$ & $(2m,-1)$ & $\ol3_{-1/2,-1/2}$ & $(0,-1)$ \\
      &          &               &           & $1_{-1/2,+3/2}$    & $(+2,-1)$ \\
\hline
$8_c$ & $(2m+1,2n+1)$ & $4_{-1/2}$ & $(2m+1,-1)$ & $3_{-1/2,+1/2}$ & $(+1,-1)$ \\
      &             &            &           & $1_{-1/2,-3/2}$ & $(-1,-1)$ \\
\cline{3-6}
      &        & $\ol4_{+1/2}$ & $(2m+1,+1)$ & $\ol3_{+1/2,-1/2}$ & $(-1,+1)$ \\
      &        &               &           & $1_{+1/2,+3/2}$    & $(+1,+1)$ \\
\hline
$1$ & $(2m,2n)$ & $1_0$ & $(2m,0)$ & $1_{0,0}$ & $(0,0)$ \\
\hline
\end{tabular}}
\end{table}

\section{$E_8$ flavour multiplet}

\begin{figure}[hbt]
\centerline{
\epsfbox{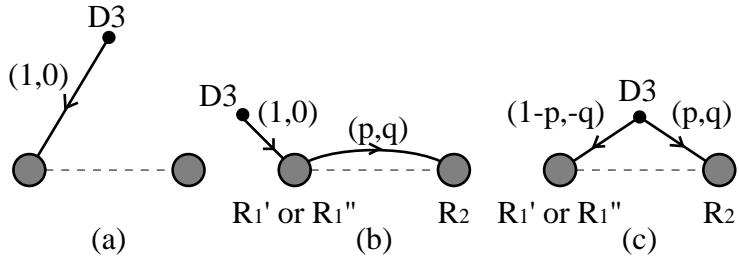}
}
\caption{The fundamental quarks correspond to D7-D3 open strings (a).
Combining D7-D3 strings and $E_8$ gauge multiplet configurations,
we obtain configurations of $E_8$ flavour multiplet (b).
This can be regarded as the bound states of two dyons (c).}
\label{fig:quarks}
\end{figure}

In terms of field theory on D3-branes,
the gauge symmetry enhancement in the previous section is interpreted
as global symmetry enhancement.
If $E_8$ symmetry is realized, quark multiplet
should be extended to $E_8$ multiplet whose dimension is at least 248.
What are these extra degrees of freedom?
Perturbative quark multiplets are supplied from open strings
stretched between the D3-brane and D7-branes in each blocks
((a) in Fig.\,\ref{fig:quarks}).
On the other hand, we know string configurations of adjoint
representation of $E_8$, which play the role of $E_8$ generators.
Therefore we get $E_8$ flavour multiplets by combining these two
configurations.
The obtained configurations are shown in (b) of Fig.\,\ref{fig:quarks}.
If we move the D3-brane into the curve of marginal stability,
which is a closed curve goes through the two blocks,
the configurations decay into
$(p,q)$ and $(1-p,-q)$ strings ((c) of Fig.\,\ref{fig:quarks}).
The $(p,q)$ strings attached on D3-branes are understood as dyons
with electric charge $p$ and magnetic charge $q$.
Therefore, we suggest that configurations (b) in Fig.\,\ref{fig:quarks}
are regarded as bound states of two dyons with charge $(p,q)$ and $(1-p,-q)$.
(Similar processes related to the splitting of orientifold planes are studied
in \cite{junction}.)
These two dyons belong to $R'_1$ of $SO(6)$ or $R''_1$ of $U(3)$
and $R_2$ of $SO(8)$, respectively. 
They combine to make bound states, which
constitute $E_8$ multiplet together with fundamental quarks.

There is a cleverer way to construct the $E_8$ flavour multiplets than
combining perturbative quark configurations and $E_8$ gauge
configurations. 
In order to obtain $E_8$ flavour multiplet, all we have to do is
setting $n_F=-1$ in configuration (a) of Fig.\ref{fig:987} or setting
$n_F=-1$ and $n_D=0$ in (b) of Fig.\ref{fig:987},
and replacing the decoupled D7-brane to a D3-brane, on which ${\cal N}=2$
$SU(2)$ Yang-Mills theory is realized.
Charges of strings stretched between two blocks in each configuration
are determined by eq.(\ref{reduction}) and eq.(\ref{eq2}).

In the case of (a) in Fig.\ref{fig:987},
the configuration contains seven D7-branes, two B-branes and two C-branes,
and the charge $(p,q)$ is given by $(p,q)=(p_2,q_2+2)$.
The shift by $+2$ of D-string charge can be regarded as contribution of
D-string loop which is added to configurations in Fig.\ref{fig:68}
to change $n_F$ by $-1$.
As a result we obtain correspondence between pairs of representations
$(R'_1,R_2)$ under global symmetry $SO(6)\times SO(8)$ and
charges of dyons $(1-p_2,-q_2-2)$ and $(p_2,q_2+2)$ (Table \ref{tbl:5dspec}).
\begin{table}[hbt]
\caption{$E_8$ flavour multiplets in five dimension
consist of bound states of two dyons.
The two dyons carry $SO(6)$ and $SO(8)$ charges of perturbative
flavour groups respectively.
The 3rd and 7th lines correspond to fundamental quarks.}
\label{tbl:5dspec}
\centerline{%
\begin{tabular}{cc|cc}
\hline
$R'_1$ & dyon charge &
$R_2$ & dyon charge \\
\hline\hline
 $15_0$        & $(-2m+1,-2)$  & $1$  & $(2m,+2)$    \\
 $6_{+1}$      & $(-2m+1,-4)$ & $1$  & $(2m,+4)$   \\
 $6_{-1}$      & $(-2m+1,0)$ & $1$  & $(2m,0)$   \\
 $1_0$         & $(-2m+1,-2)$  & $1$  & $(2m,+2)$    \\
 $6_0$         & $(-2m,-2)$    & $8_v$ & $(2m+1,+2)$  \\
 $1_{+1}$      & $(-2m,-4)$   & $8_v$ & $(2m+1,+4)$ \\
 $1_{-1}$      & $(-2m,0)$   & $8_v$ & $(2m+1,0)$ \\
 $4_{+1/2}$    & $(-2m+1,-3)$ & $8_s$ & $(2m,+3)$   \\
 $\ol4_{-1/2}$ & $(-2m+1,-1)$ & $8_s$ & $(2m,+1)$   \\
 $4_{-1/2}$    & $(-2m,-1)$   & $8_c$ & $(2m+1,+1)$ \\
 $\ol4_{+1/2}$ & $(-2m,-3)$   & $8_c$ & $(2m+1,+3)$ \\
 $1_0$ & $(-2m+1,-2)$ & $28$ & $(2m,+2)$ \\
\hline
\end{tabular}}
\end{table}
The fundamental quarks correspond to the 3rd and 7th line
in Table \ref{tbl:5dspec}.
In this case, string charge of dyons have mod $2$ ambiguity
corresponding to fundamental string loops.
Because fundamental string loops are transformed into Kaluza-Klein momentum
via T-duality, the field theory on D3-brane is regarded as five dimensional
$SU(2)$ Yang-Mills theory on D4-brane, which is T-dual of the D3-brane.

If we use configuration (b) in Fig.\ref{fig:987},
in which one D7-brane and one B-brane are decoupled, string charge $(p,q)$,
which is determined by eq.(\ref{reduction}) and eq.(\ref{eq2}),
is given by $(p,q)=(p_3,q_3+2)$ and we obtain spectrum
shown in Table \ref{tbl:68}.
\begin{table}[hbt]
\caption{$E_8$ flavour multiplets in four dimension consist of
bound states of two dyons.
The two dyons carry $SU(3)$ and $SO(8)$ charges of perturbative
flavour groups respectively.
The 7th and 13th lines correspond to fundamental quarks.}
\label{tbl:68}
\centerline{%
\begin{tabular}{cc|cc}
\hline
$R''_1$ & dyon charge &
$R_2$ & dyon charge \\
\hline\hline
 $8_{0,0}$      & $(+1,-2)$ & $1$ & $(0,+2)$ \\
 $3_{0,+2}$     & $(-1,-2)$ & $1$ & $(+2,+2)$ \\
 $\ol3_{0,-2}$  & $(+3,-2)$ & $1$ & $(-2,+2)$ \\
 $1_{0,0}$      & $(+1,-2)$ & $1$ & $(0,+2)$ \\
 $3_{+1,-1}$    & $(+3,-4)$ & $1$ & $(-2,+4)$ \\
 $\ol3_{+1,+1}$ & $(+1,-4)$ & $1$ & $(0,+4)$ \\
 $3_{-1,-1}$    & $(+1,0)$  & $1$ & $(0,0)$ \\
 $\ol3_{-1,+1}$ & $(-1,0)$  & $1$ & $(+2,0)$ \\
 $1_{0,0}$      & $(+1,-2)$ & $1$ & $(0,+2)$ \\
 $3_{0,-1}$     & $(+2,-2)$ & $8_v$ & $(-1,+2)$ \\
 $\ol3_{0,+1}$  & $(0,-2)$  & $8_v$ & $(+1,+2)$ \\
 $1_{+1,0}$     & $(+2,-4)$ & $8_v$ & $(-1,+4)$ \\
 $1_{-1,0}$     & $(0,0)$   & $8_v$ & $(+1,0)$ \\
 $3_{+1/2,+1/2}$    & $(+1,-3)$ & $8_s$ & $(0,+3)$ \\
 $1_{+1/2,-3/2}$    & $(+3,-3)$ & $8_s$ & $(-2,+3)$ \\
 $\ol3_{-1/2,-1/2}$ & $(+1,-1)$ & $8_s$ & $(0,+1)$ \\
 $1_{-1/2,+3/2}$    & $(-1,-1)$ & $8_s$ & $(+2,+1)$ \\
 $3_{-1/2,+1/2}$    & $(0,-1)$  & $8_c$ & $(+1,+1)$ \\
 $1_{-1/2,-3/2}$    & $(+2,-1)$ & $8_c$ & $(-1,+1)$ \\
 $\ol3_{+1/2,-1/2}$ & $(+2,-3)$ & $8_c$ & $(-1,+3)$ \\
 $1_{+1/2,+3/2}$    & $(0,-3)$  & $8_c$ & $(+1,+3)$ \\
 $1_{0,0}$          & $(+1,-2)$ & $28$  & $(0,+2)$ \\
\hline
\end{tabular}}
\end{table}
The fundamental quarks correspond to the 7th and 13th lines
in Table \ref{tbl:68}.
In this case, both of string charge $p$ and $q$ are fixed.
Therefore, these configurations give $E_8$ flavour multiplet of
four dimensional $SU(2)$ Yang-Mills theory on the D3-brane.

\section{Conclusions}

We have presented string configurations which belong to adjoint, vector, and
spinor representations of $SO(8)$.
Combining them, we constructed configurations in $\bf248$ of $E_8$
which live on nine spatial dimension.
We showed that the decoupling of one D7-brane and one B-brane
does not break gauge symmetry
but decreases the space-time dimension where the $E_8$ gauge field lives.
This is explained by means of possibility of adding string loops.
If we decouple one D7-brane, the dimension decreases by one
and the field theory is regarded as $E_8$ gauge theory on D8-brane.
Further decoupling of a B-brane gives seven spatial dimensional
field theory and its string configurations
are already given in \cite{FromOpen}.
Our construction based on $SO(8)$ decomposition
is convenient to see the relationships to type-I' theory.
In fact, we showed that the spectrum
agrees completely with the result in \cite{Creation}.
Once we decouple both a D7-brane and a B-brane,
the ambiguity of string charge is fixed.
Therefore, further decouplings of D7-branes break gauge symmetry
and configurations for $E_7$, $E_6$,$\ldots$ will be obtained.
Using these configurations, we constructed string configurations
of $E_8$ flavour multiplet in five and four dimensions.
These configurations can be interpreted as bound states of
two dyons.

\vspace{1ex}

\noindent{\bf Acknowledgment}

I would like to thank H.\ Hata
for careful reading of the manuscript.


\end{document}